\documentclass[a4paper, oribibl]{llncs}  % With Running heads

\usepackage{microtype, hyphenat, amssymb, amsmath, cite}
\usepackage{graphicx, wrapfig}
\usepackage{hyperref}
\usepackage{cleveref}
\usepackage{setspace}
\usepackage[figuresleft]{rotating}
\usepackage{url}
\newcommand{\keywords}[1]{\par\addvspace\baselineskip
\noindent\keywordname\enspace\ignorespaces#1}

\begin{document}
\mainmatter  % start of an individual contribution
\title{On Vulnerabilities of the Security Association in the IEEE 802.15.6 Standard
\thanks{Note: A variant of this paper will appear in \cite{T15b}.}}
\titlerunning{On Vulnerabilities of the Security Association in the IEEE 802.15.6 Standard}  
\author{Mohsen Toorani}
\authorrunning{M. Toorani}
\institute{Department of Informatics\\University of Bergen, Norway\\ %\email{mohsen.toorani@ii.uib.no}
}
\toctitle{On Vulnerabilities of the Security Association in the IEEE 802.15.6 Standard}
\tocauthor{Mohsen Toorani}
\maketitle

\begin{abstract}
Wireless Body Area Networks (WBAN) support a variety of real-time health monitoring and consumer electronics applications. The latest international standard for WBAN is the IEEE 802.15.6. The security association in this standard includes four elliptic curve-based key agreement protocols that are used for generating a master key. In this paper, we challenge the security of the IEEE 802.15.6 standard by showing vulnerabilities of those four protocols to several attacks. We perform a security analysis on the protocols, and show that they all have security problems, and are vulnerable to different attacks.
\end{abstract}

\section{Introduction}
\label{sec:1}
Advances in wireless communication and embedded computing technologies, such as wearable and implantable biosensors, have enabled the design, development, and implementation of \emph{Body Area Networks} (BAN) \cite{chen2011body}. A BAN, also referred to as a \emph{Wireless Body Area Network} (WBAN) or a \emph{Body Sensor Network} (BSN), is a wireless network of wearable computing devices. BAN devices may be embedded inside the body (implants), may be mounted on the body (wearable technology), or may be accompanying devices that humans can carry in clothes pockets, by hand or in various bags. WBAN can be used for many applications such as military, ubiquitous health care, sport, and entertainment \cite{WBAN14survey, chen2011body}. WBANs have a huge potential to revolutionize the future of health care monitoring by diagnosing many life-threatening diseases, and providing real-time patient monitoring \cite{WBAN14survey}. WBANs may interact with the Internet and other existing wireless technologies.

The latest standardization of WBANs is the IEEE 802.15.6 standard \cite{WBANstandard} which aims to provide an international standard for low power, short range, and extremely reliable wireless communication within the surrounding area of the human body, supporting a vast range of data rates for different applications.

The network topology consists of nodes and hubs. A node is an entity that contains a Medium Access Control (MAC) sublayer and a physical (PHY) layer, and optionally provides security services. A hub is an entity that possesses a node's functionality, and coordinates the medium access and power management of the nodes. Nodes can be classified into different groups based on their functionality (personal devices, sensors, actuators), implementation (implant nodes, body surface nodes, external nodes) and role (coordinators, end nodes, relays) \cite{WBAN14survey}.

Although security is a high priority in most networks, little study has been done in this area for WBANs. As WBANS are resource-constrained in terms of power, memory, communication rate and computational capability, security solutions proposed for other networks may not be applicable to WBANs. Confidentiality, authentication, integrity, and freshness of data together with availability and secure management are the security requirements in WBAN \cite{WBAN14survey}. A \emph{security association} is a procedure in the IEEE 802.15.6 standard to identify a node and a hub to each other, to establish a new \emph{Master Key} ($MK$) shared between them, or to activate an existing $MK$ pre-shared between them. The security association in the IEEE 802.15.6 standard is based on four key agreement protocols that are presented in the standard \cite{WBANstandard}.

Authenticated Key Exchange (AKE) and Password-Authenticated Key Exchange (PAKE) protocols aim to exchange a cryptographic session key between legitimate entities in an authenticated manner. Several security properties must be satisfied by AKE and PAKE protocols, and they should obviously withstand well-known attacks. Many protocols have been proposed in the literature, but some of them have been shown to have security problems \cite{HMQV, menezes07another, T15a}. It is desirable for AKE protocols to provide known-key security, forward secrecy, key control, and resilience to well-known attacks such as Key-Compromise Impersonation (KCI) and its variants, unknown key-share (UKS), replay, and Denning-Sacco attacks. %\cite{T14a, MQV, HMQV, eCK}.
PAKE protocols must also be resilient to dictionary attacks \cite{T14a, BPR}.

In this paper, we perform a security analysis on four key agreements protocols that are used in the security association process of the IEEE 802.15.6 standard \cite{WBANstandard}. We challenge the security of the IEEE 802.15.6 standard by showing vulnerabilities of those four protocols to several attacks. Excluding vulnerability of the first protocol to the impersonation attack which has been implied in the standard, no attack or security vulnerability has been reported in the standard or literature. All the protocols are available in the latest version of the IEEE 802.15.6 standard.
The rest of this paper is organized as follows. We review the security structure of the IEEE 802.15.6 standard in Section \ref{sec:2}, these key agreement protocols in Section \ref{sec:3}, and report their security problems in Section \ref{sec:4}.

\section{Security Structure of the IEEE 802.15.6 standard}
\label{sec:2}
\begin{figure} [!t]
  \centering
  \includegraphics[width=12cm]{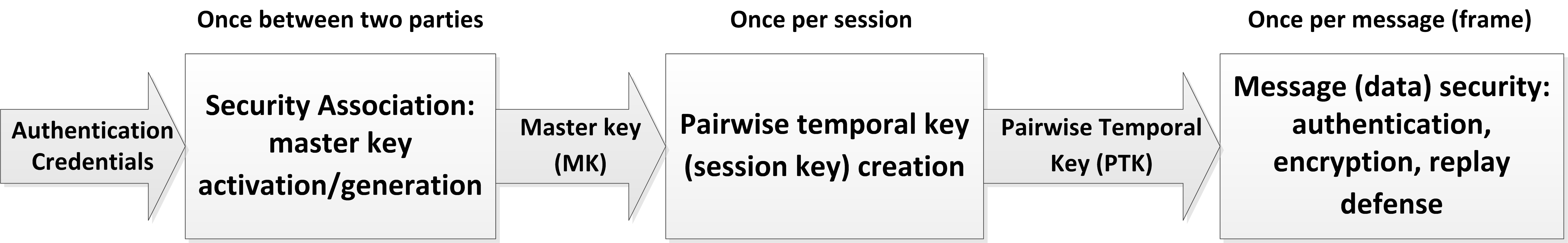}\\
  \caption{Security hierarchy in IEEE 802.15.6 Standard \cite{WBANstandard}}
  \label{fig:01}
\end{figure}

The Security hierarchy of the IEEE 802.15.6 standard is depicted in Figure \ref{fig:01}. All nodes and hubs must choose three security levels: unsecured communication (level 0), authentication but no encryption (level 1), and authentication and encryption (level 2). During the security association process, a node and a hub need to jointly select a suitable security level. In unicast communication, a pre-shared or a new $MK$ is activated. A \emph{Pairwise Temporal Key} ($PTK$) is then generated that is used only once per session. In multicast communication, a \emph{Group Temporal Key} ($GTK$) is generated that is shared with the corresponding group \cite{WBANstandard}.
All nodes and hubs in a WBAN have to go through certain stages at the MAC layer before data exchange. The security state diagrams of the IEEE 802.15.6 Standard for secured and unsecured communication are depicted in Figure \ref{fig:02}. In a secured communication, a node can be in one of following states \cite{WBANstandard}:
\begin{figure} [!t]
  \centering
  \includegraphics[width=12cm]{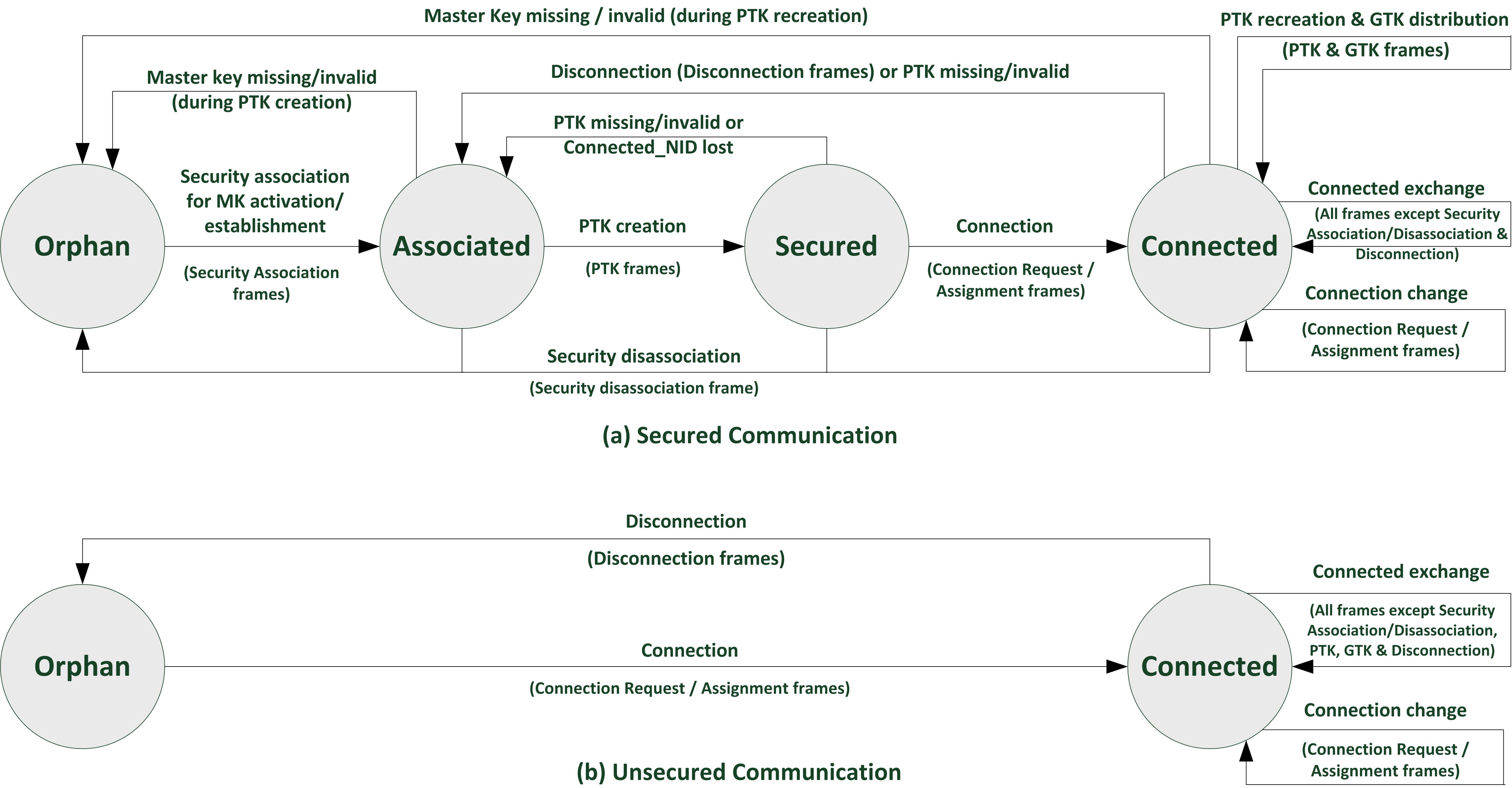}\\
  \caption{MAC and security state diagrams in IEEE 802.15.6 Standard \cite{WBANstandard}}
  \label{fig:02}
\end{figure}

\begin{itemize}
  \item \emph{Orphan}: The initial state where the node does not have any relationship with the hub for secured communication. The node should activate a pre-shared $MK$ or share a new $MK$ with the hub. They cannot proceed to the Associated state if they fail to activate/establish a shared $MK$.
  \item \emph{Associated}: The node holds a shared $MK$ with the hub for their $PTK$ creation. The node and hub are allowed to exchange $PTK$ frames with each other to confirm the possession of a shared $MK$, create a $PTK$ and transit to the Secured state. If the $MK$ is invalid/missing during the $PTK$ creation, they will move back to the Orphan state.
  \item \emph{Secured}: The node holds a $PTK$ with the hub.
  The node and hub can exchange security disassociation frames, connection assignment secure frames, connection request and control unsecured frames. The node can exchange Connection Request and Connection Assignment frames with the hub to form a connection, and transit to the Connected state.
  \item \emph{Connected}: The node holds an assigned \emph{Connected\_NID}, a wakeup arrangement, and optionally one or more scheduled and unscheduled allocations with the hub for abbreviated node addressing, desired wakeup, and optionally scheduled and unscheduled access. The node and hub are not allowed to send any unsecured frame to each other, other than unsecured control frames if authentication of control type frames was not selected during the association.
\end{itemize}

\section{Key Agreement Protocols in the IEEE 802.15.6 Standard}
\label{sec:3}

\begin{sidewaysfigure*}
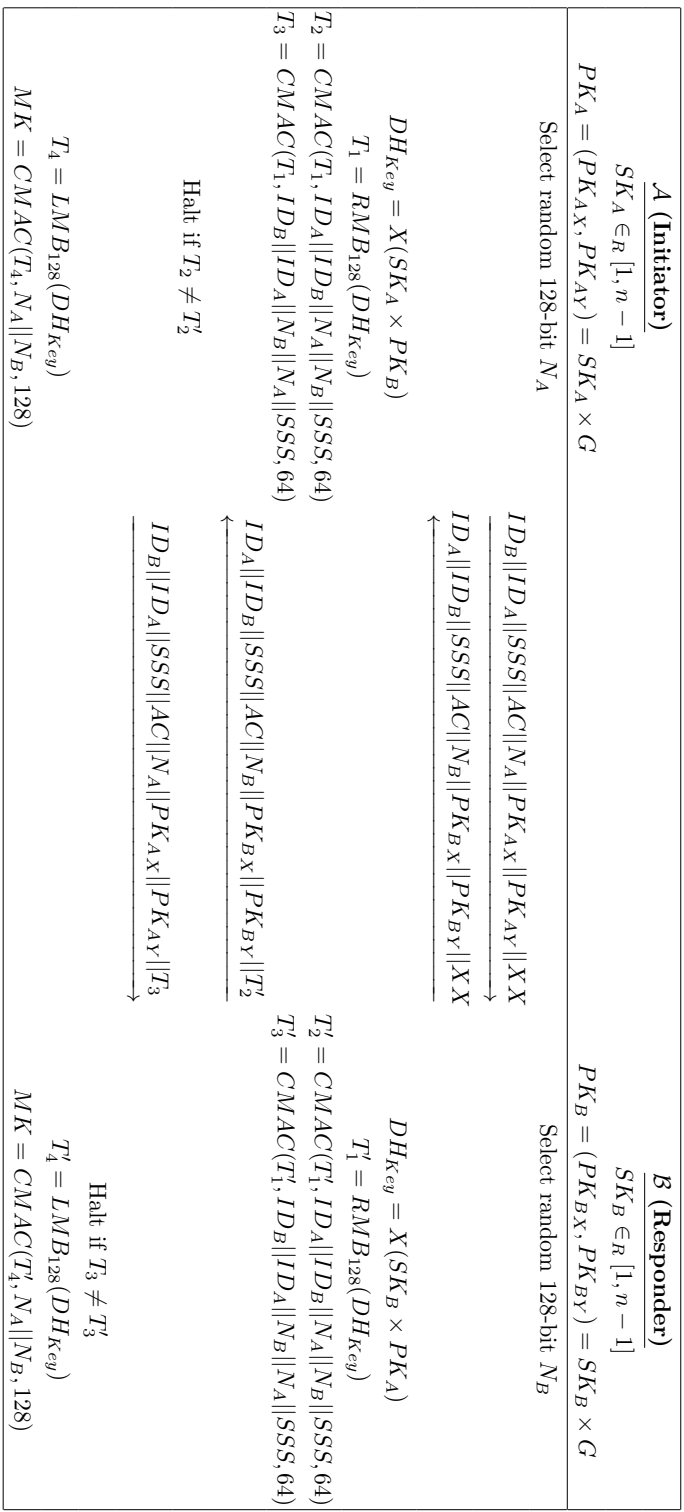
 %[!t]
\begin{center}
\scalebox{.9}{
\begin{tabular}{|c c c|}
\hline
\underline{\bf{$\mathcal{A}$ (Initiator)}}&  &\underline{\bf{$\mathcal{B}$ (Responder)}}\\
$SK_A \in_R [1, n-1]$ && $SK_B \in_R [1, n-1]$ \\
$PK_A = ({PK_A}_X, {PK_A}_Y) = SK_A \times G$    &&  $PK_B =  ({PK_B}_X, {PK_B}_Y) = SK_B \times G$\\
\hline
Select random 128-bit $N_A$ && Select random 128-bit $N_B$ \\
                      & $ \overset{\mbox{$ID_B || ID_A || SSS || AC || N_A || {PK_A}_X || {PK_A}_Y || XX$}}{\xrightarrow{\hspace*{7cm}}}$  &  \\
                      & $ \overset{\mbox{$ID_A || ID_B || SSS || AC || N_B || {PK_B}_X || {PK_B}_Y || XX$}}{\xleftarrow {\hspace*{7cm}}}$  &  \\
$DH_{Key} = X(SK_A \times PK_B)$ && $DH_{Key} = X(SK_B \times PK_A)$ \\
$T_1 = RMB_{128}(DH_{Key})$ &&  $T'_1 = RMB_{128}(DH_{Key})$ \\
$T_2 = CMAC(T_1, ID_A || ID_B || N_A || N_B || SSS, 64)$ &&  $T'_2 = CMAC(T'_1, ID_A || ID_B || N_A || N_B || SSS, 64)$ \\
$T_3 = CMAC(T_1, ID_B || ID_A || N_B || N_A || SSS, 64)$ &&  $T'_3 = CMAC(T'_1, ID_B || ID_A || N_B || N_A || SSS, 64)$ \\
                      & $ \overset{\mbox{$ID_A || ID_B || SSS || AC || N_B || {PK_B}_X || {PK_B}_Y || T'_2$}}{\xleftarrow {\hspace*{7cm}}}$  &  \\
Halt if $T_2 \neq T'_2$ && \\
                      & $ \overset{\mbox{$ID_B || ID_A || SSS || AC || N_A || {PK_A}_X || {PK_A}_Y || T_3$}}{\xrightarrow{\hspace*{7cm}}}$  &  \\
&& Halt if $T_3 \neq T'_3$ \\
$T_4 = LMB_{128}(DH_{Key})$ && $T'_4 = LMB_{128}(DH_{Key})$ \\
$MK = CMAC(T_4, N_A || N_B, 128)$ && $MK = CMAC(T'_4, N_A || N_B, 128)$ \\
\hline
\end{tabular}
}
\caption{Unauthenticated key agreement protocol (Protocol I). }
\label{fig:1}
\end{center}
\end{sidewaysfigure*}

\begin{sidewaysfigure*}
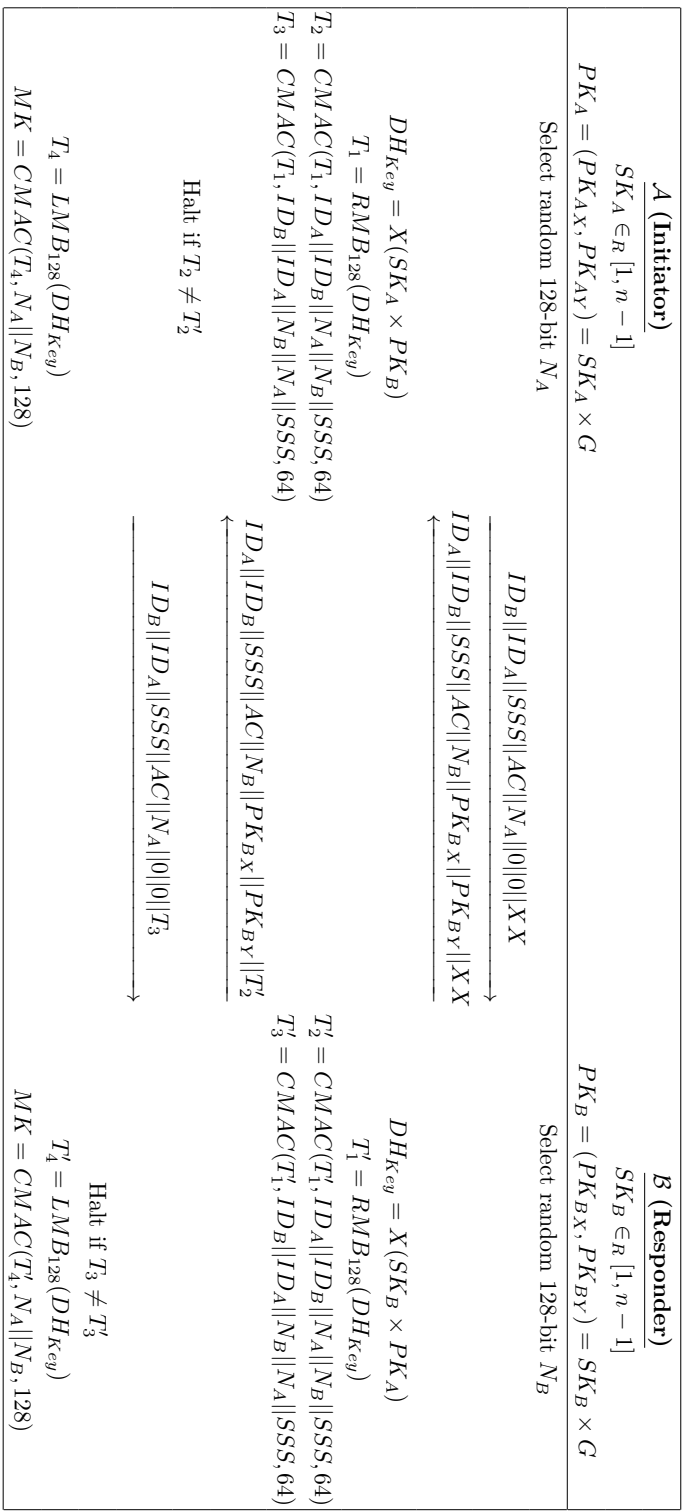
 %[!t] %[htbp]
\begin{center}
\scalebox{.9}{
\begin{tabular}{|c c c|}
\hline
\underline{\bf{$\mathcal{A}$ (Initiator)}}&  &\underline{\bf{$\mathcal{B}$ (Responder)}}\\
$SK_A \in_R [1, n-1]$ && $SK_B \in_R [1, n-1]$ \\
$PK_A = ({PK_A}_X, {PK_A}_Y) = SK_A \times G$    &&  $PK_B =  ({PK_B}_X, {PK_B}_Y) = SK_B \times G$\\
\hline
Select random 128-bit $N_A$ && Select random 128-bit $N_B$ \\
                      & $ \overset{\mbox{$ID_B || ID_A || SSS || AC || N_A || 0 || 0 || XX$}}{\xrightarrow{\hspace*{7cm}}}$  &  \\
                      & $ \overset{\mbox{$ID_A || ID_B || SSS || AC || N_B || {PK_B}_X || {PK_B}_Y || XX$}}{\xleftarrow {\hspace*{7cm}}}$  &  \\
$DH_{Key} = X(SK_A \times PK_B)$ && $DH_{Key} = X(SK_B \times PK_A)$ \\
$T_1 = RMB_{128}(DH_{Key})$ &&  $T'_1 = RMB_{128}(DH_{Key})$ \\
$T_2 = CMAC(T_1, ID_A || ID_B || N_A || N_B || SSS, 64)$ &&  $T'_2 = CMAC(T'_1, ID_A || ID_B || N_A || N_B || SSS, 64)$ \\
$T_3 = CMAC(T_1, ID_B || ID_A || N_B || N_A || SSS, 64)$ &&  $T'_3 = CMAC(T'_1, ID_B || ID_A || N_B || N_A || SSS, 64)$ \\
                      & $ \overset{\mbox{$ID_A || ID_B || SSS || AC || N_B || {PK_B}_X || {PK_B}_Y || T'_2$}}{\xleftarrow {\hspace*{7cm}}}$  &  \\
Halt if $T_2 \neq T'_2$ && \\
                      & $ \overset{\mbox{$ID_B || ID_A || SSS || AC || N_A || 0 || 0 || T_3$}}{\xrightarrow{\hspace*{7cm}}}$  &  \\
&& Halt if $T_3 \neq T'_3$ \\
$T_4 = LMB_{128}(DH_{Key})$ && $T'_4 = LMB_{128}(DH_{Key})$ \\
$MK = CMAC(T_4, N_A || N_B, 128)$ && $MK = CMAC(T'_4, N_A || N_B, 128)$ \\
\hline
\end{tabular}
}
\caption{Hidden public key transfer authenticated key agreement protocol (Protocol II). }
\label{fig:2}
\end{center}
\end{sidewaysfigure*}

\begin{sidewaysfigure*} %[!t]
\begin{center}
\scalebox{.9}{
\begin{tabular}{|c c c|}
\hline
\underline{\bf{$\mathcal{A}$ (Initiator)}}&  &\underline{\bf{$\mathcal{B}$ (Responder)}}\\
$SK_A \in_R [1, n-1]$ && $SK_B \in_R [1, n-1]$ \\
$PK_A = ({PK_A}_X, {PK_A}_Y) = SK_A \times G$    &&  $PK_B =  ({PK_B}_X, {PK_B}_Y) = SK_B \times G$\\
$PK'_A = PK_A - Q(PW)$ && \\
Update $SK_A$ and $PK_A$ if $PK'_A = O$ && \\
\hline
Select random 128-bit $N_A$ && Select random 128-bit $N_B$ \\
                      & $ \overset{\mbox{$ID_B || ID_A || SSS || AC || N_A || {PK'_A}_X || {PK'_A}_Y || XX$}}{\xrightarrow{\hspace*{7cm}}}$  &  \\
                      & $ \overset{\mbox{$ID_A || ID_B || SSS || AC || N_B || {PK_B}_X || {PK_B}_Y || XX$}}{\xleftarrow {\hspace*{7cm}}}$  &  \\
&& $PK_A = PK'_A + Q(PW)$ \\
$DH_{Key} = X(SK_A \times PK_B)$ && $DH_{Key} = X(SK_B \times PK_A)$ \\
$T_1 = RMB_{128}(DH_{Key})$ &&  $T'_1 = RMB_{128}(DH_{Key})$ \\
$T_2 = CMAC(T_1, ID_A || ID_B || N_A || N_B || SSS, 64)$ &&  $T'_2 = CMAC(T'_1, ID_A || ID_B || N_A || N_B || SSS, 64)$ \\
$T_3 = CMAC(T_1, ID_B || ID_A || N_B || N_A || SSS, 64)$ &&  $T'_3 = CMAC(T'_1, ID_B || ID_A || N_B || N_A || SSS, 64)$ \\
                      & $ \overset{\mbox{$ID_A || ID_B || SSS || AC || N_B || {PK_B}_X || {PK_B}_Y || T'_2$}}{\xleftarrow {\hspace*{7cm}}}$  &  \\
Halt if $T_2 \neq T'_2$ && \\
                      & $ \overset{\mbox{$ID_B || ID_A || SSS || AC || N_A || {PK_A}_X || {PK_A}_Y || T_3$}}{\xrightarrow{\hspace*{7cm}}}$  &  \\
&& Halt if $T_3 \neq T'_3$ \\
$T_4 = LMB_{128}(DH_{Key})$ && $T'_4 = LMB_{128}(DH_{Key})$ \\
$MK = CMAC(T_4, N_A || N_B, 128)$ && $MK = CMAC(T'_4, N_A || N_B, 128)$ \\
\hline
\end{tabular}
}
\caption{Password authenticated association procedure (Protocol III). }
\label{fig:3}
\end{center}
\end{sidewaysfigure*}

\begin{sidewaysfigure*} [!htbp]
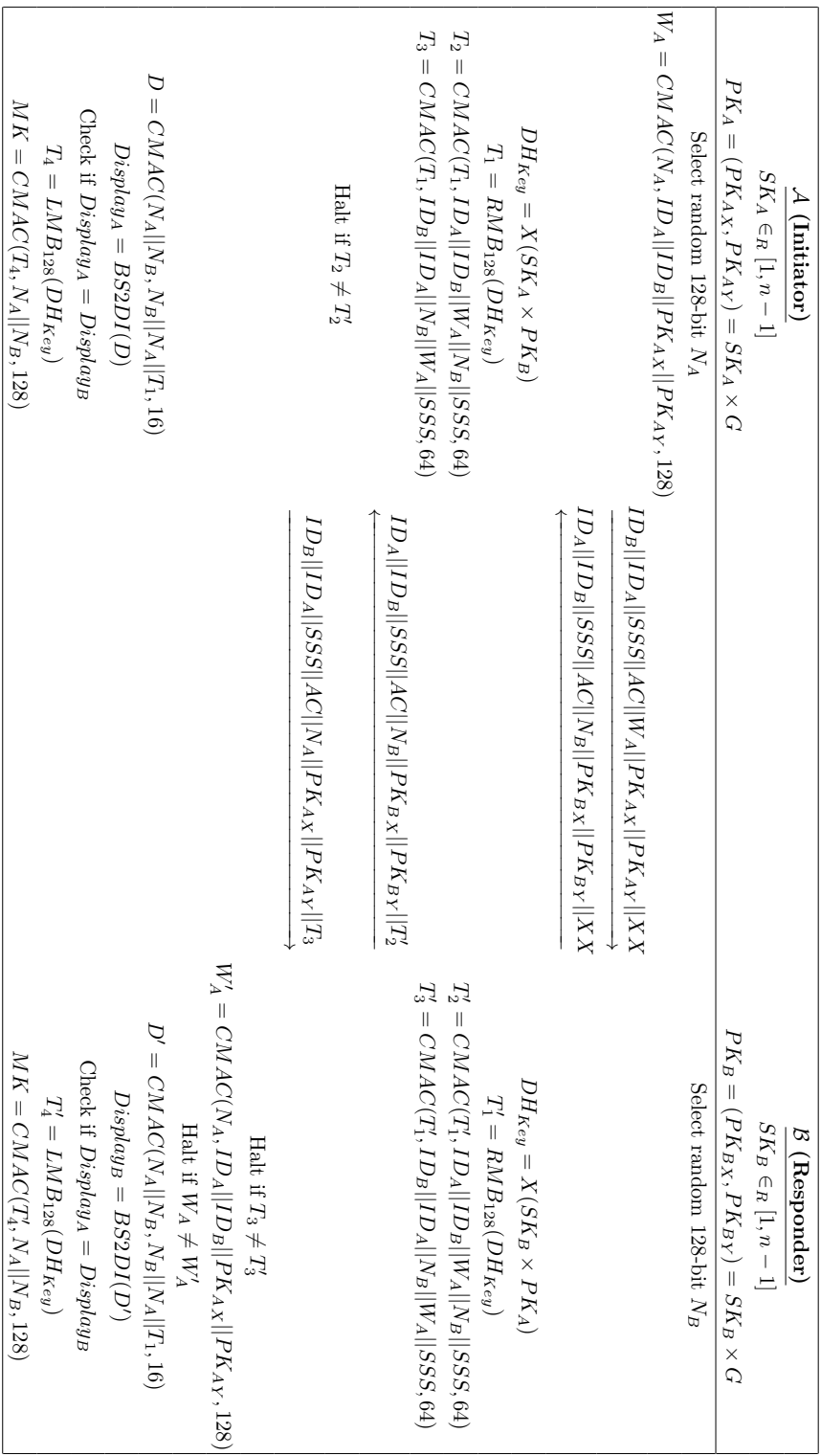

%\centering
\begin{center}
\scalebox{.85}{
\begin{tabular}{|c c c|}
\hline
\underline{\bf{$\mathcal{A}$ (Initiator)}}&  &\underline{\bf{$\mathcal{B}$ (Responder)}}\\
$SK_A \in_R [1, n-1]$ && $SK_B \in_R [1, n-1]$ \\
$PK_A = ({PK_A}_X, {PK_A}_Y) = SK_A \times G$    &&  $PK_B =  ({PK_B}_X, {PK_B}_Y) = SK_B \times G$\\
\hline
Select random 128-bit $N_A$ && Select random 128-bit $N_B$ \\
$W_A = CMAC(N_A, ID_A || ID_B || {PK_A}_X || {PK_A}_Y, 128)$ && \\
                      & $ \overset{\mbox{$ID_B || ID_A || SSS || AC || W_A || {PK_A}_X || {PK_A}_Y || XX$}}{\xrightarrow{\hspace*{7cm}}}$  &  \\
                      & $ \overset{\mbox{$ID_A || ID_B || SSS || AC || N_B || {PK_B}_X || {PK_B}_Y || XX$}}{\xleftarrow {\hspace*{7cm}}}$  &  \\
$DH_{Key} = X(SK_A \times PK_B)$ && $DH_{Key} = X(SK_B \times PK_A)$ \\
$T_1 = RMB_{128}(DH_{Key})$ &&  $T'_1 = RMB_{128}(DH_{Key})$ \\
$T_2 = CMAC(T_1, ID_A || ID_B || W_A || N_B || SSS, 64)$ &&  $T'_2 = CMAC(T'_1, ID_A || ID_B || W_A || N_B || SSS, 64)$ \\
$T_3 = CMAC(T_1, ID_B || ID_A || N_B || W_A || SSS, 64)$ &&  $T'_3 = CMAC(T'_1, ID_B || ID_A || N_B || W_A || SSS, 64)$ \\
                      & $ \overset{\mbox{$ID_A || ID_B || SSS || AC || N_B || {PK_B}_X || {PK_B}_Y || T'_2$}}{\xleftarrow {\hspace*{7cm}}}$  &  \\
Halt if $T_2 \neq T'_2$ && \\
                      & $ \overset{\mbox{$ID_B || ID_A || SSS || AC || N_A || {PK_A}_X || {PK_A}_Y || T_3$}}{\xrightarrow{\hspace*{7cm}}}$  &  \\
&& Halt if $T_3 \neq T'_3$ \\
&& $W'_A = CMAC(N_A, ID_A || ID_B || {PK_A}_X || {PK_A}_Y, 128)$ \\
&& Halt if $W_A \neq W'_A$ \\
$D = CMAC(N_A || N_B, N_B|| N_A || T_1, 16)$ && $D' = CMAC(N_A || N_B, N_B|| N_A || T_1, 16)$ \\
$Display_A = BS2DI(D)$ && $Display_B = BS2DI(D')$ \\
Check if $Display_A = Display_B$  &&  Check if $Display_A = Display_B$ \\
$T_4 = LMB_{128}(DH_{Key})$ && $T'_4 = LMB_{128}(DH_{Key})$ \\
$MK = CMAC(T_4, N_A || N_B, 128)$ && $MK = CMAC(T'_4, N_A || N_B, 128)$ \\
\hline
\end{tabular}
}
\caption{Display authenticated association procedure (Protocol IV). }
\label{fig:4}
\end{center}
\end{sidewaysfigure*}

The security association in the 802.15.6 standard involves a Master Key ($MK$) which is generated using one of four two-party key agreement protocols, proposed in the standard. Those four protocols, that will be referred to as protocols I-IV in this paper, are depicted in \Cref{fig:1,fig:2,fig:3,fig:4}. The goal is to establish a new $MK$ between a node and a hub. The node and hub are denoted by $\mathcal{A}$ and $\mathcal{B}$, respectively. For simplicity, we have used simpler notations than those of the standard \cite{WBANstandard}. We have deleted \emph{Immediate Acknowledgement} (I-Ack) messages that $\mathcal{B}$ sends to $\mathcal{A}$, after receiving each frame from $\mathcal{A}$.
I-Ack is kind of control type frames, and consists of current allocation slot number (8 bits) and current allocation slot offset (16 bits). We deleted I-Ack because they are sent in clear. Any information sent in clear, can be deleted from the security analysis.
Protocols I-IV are similar, but vary in details and requirements.
Protocol I is unauthenticated, and does not have any special requirement. Protocol II requires pre-shared and out-of-band transfer of a node's public key to the hub. Then, it is assumed that a hub obtains a node's public key via a separate protected channel, and a hub needs to save public keys of the nodes. Protocol III requires that a node and hub pre-share a password ($PW$). Protocol IV requires that $\mathcal{A}$ and $\mathcal{B}$ each has a display that shows a decimal number. It also requires that before accepting a new $MK$, human user(s) verify that both displays show the same number.

Protocols I-IV are based on elliptic curve public key cryptography. The domain parameters consist of an elliptic curve with Weierstrass equation of the form $y^2 = x^3 + ax + b$, defined over the finite field $GF(p)$ where $p$ is a prime number. In order to make the elliptic curve non-singular, $a,b \in GF(p)$ should satisfy $4a^3 + 27b^2 \neq0$. The base point $G$ in the elliptic curve is of order $n$, where $n \times G = O$ in which $O$ denotes the point at infinity. The IEEE 802.15.6 standard suggests using the Curve P-256 in FIPS Pub 186-3. Values of $a, b, p, n$ and $G$ are public, and given in \cite{WBANstandard}.

The private keys shall be 256-bit random integers, chosen independently from the set of integers $\{1,...,n-1\}$. The private key of $\mathcal{A}$ and $\mathcal{B}$ is denoted by $SK_A$ and $SK_B$, respectively. The corresponding public keys are generated as $PK_A = ({PK_A}_X , {PK_A}_Y) = SK_A \times G$ and $PK_B = ({PK_B}_X , {PK_B}_Y) = SK_B \times G$.

The IEEE 802.15.6 standard does not include having a digital certificate for public keys. Public keys are self-generated by involved parties, and are not accompanied by digital certificates. It is because nodes are likely to be severely resource-constrained, and hence cannot store certificates or perform the certificate validation. The process of certificate validation consists of verifying the integrity and authenticity of the certificate by verifying the certificate authority's signature on the certificate, verifying that the certificate is not expired, and verifying that the certificate is not revoked \cite{T09b}. %\cite{stinson06cryptography}.

The standard specifies that the node and the hub will abort execution of the protocols if the received public key, sent from the other party, is not a valid public key. A received public key $PK_i = ({PK_i}_X , {PK_i}_Y)$ shall be treated valid only if it is a non-infinity point (i.e. $PK_i \neq O$) on the defined elliptic curve, i.e. ${PK_i}_X$ and ${PK_i}_Y$ satisfy the elliptic curve equation given above. This has been explained in protocol descriptions in the IEEE 802.15.6 standard, but they are absent in the corresponding figures of the standard. We do not show such verifications in \Cref{fig:1,fig:2,fig:3,fig:4} either. It is noteworthy that validation of elliptic curve public keys includes more steps than those mentioned in the standard. In addition to those conditions, one should check that ${PK_i}_X$ and ${PK_i}_Y$ are properly represented elements in $GF(p)$, and that $n \times PK_i = O$. The last condition is implied by the other three conditions if the cofactor of the elliptic curve $h = 1$, which is the case for curves over prime finite fields \cite{hankerson04guide}.

In protocols I-IV, $\mathcal{B}$ always sends his public key $PK_B$ in clear. In Protocols I and IV, $\mathcal{A}$ sends her public key in clear. In Protocols II, $\mathcal{A}$ does not send her public key, as $PK_A$ is pre-shared with $\mathcal{B}$. However, in protocol III, $\mathcal{A}$ first sends a masked public key $PK'_A = PK_A - Q(PW)$ in which $PW$ is a positive integer, converted from the pre-shared password between $\mathcal{A}$ and $\mathcal{B}$. $PW$ is converted according to the IEEE 1363-2000 standard from the UTE-16BE representation, specified in ISO/IEC 10646:2003, by treating the leftmost octet as the octet containing the \emph{Most Significant Bits} (MSB). The $Q(.)$ function is a mapping which converts the integer $PW$ to the point $Q(PW)=(Q_X,Q_Y)$ on the elliptic curve in which $Q_X = 2^{32} PW + M_X$ where $M_X$ is the smallest nonnegative integer such that $Q_X$ becomes the X-coordinate of a point on the elliptic curve. $Q_Y$ is an even positive integer, and is the Y-coordinate of that point. In protocol III, $\mathcal{A}$ shall choose a private key $SK_A$ such that the X-coordinate of $PK_A$ is not equal to the X-coordinate of $Q(PW)$.

$CMAC(K, M, L)$ represents the $L$-bit output of the \emph{Cipher-based Message Authentication Code} (CMAC), applied under key $K$ to message $M$. The standard suggests to use CMAC with the AES forward cipher function as specified in the NIST SP800-38B, and to use a 128-bit key as specified in FIPS197. $LMB_L(S)$ and $RMB_L(S)$ designates the $L$ leftmost and the $L$ rightmost bits of the bit string $S$, respectively. $X(P)$ denotes the X-coordinate of point $P$ on the elliptic curve, i.e. $X(P) = X(P_X, P_Y) = P_X$. The sign $||$ denotes concatenation of bit strings that are converted according to the IEEE 1363-2000 standard. $BS2DI(BS)$ converts the bit string $BS$ to a positive decimal integer for display.
$SSS$ is the \emph{Security\_Suite\_Selector} (16 bits), $AC$ is the \emph{Association\_Control} (16 bits), and $XX$ is $X0000000000000000$.
\emph{Security\_Suite\_Selector} specifies type of cryptographic algorithms and protocols that will be used during the protocol execution. It consists of the type of security association protocol (3 bits), i.e. binary representation of the protocol type according to our numbering I-IV, security level (2 bits), Control Frame Authentication (1 bit), cipher function (4 bits), and 6 bits reserved for future uses. \emph{Association\_Control} consists of Association Sequence Number (4 bits), Association Status (4 bits), and 8 bits reserved for future.
$SSS$ is fixed during a protocol execution, but $AC$ will be different for each message. This is because $AC$ includes the Association Sequence Number which is increased by one after each frame is sent during a protocol execution. Excluding I-Ack frames that are deleted from the protocols, there are four paths between $\mathcal{A}$ and $\mathcal{B}$ in all protocols.

\section{Security Problems}
\label{sec:4}
In this section, we show that protocols I-IV are vulnerable to several attacks. All the protocols are vulnerable to the KCI attack, and they do not provide the forward secrecy. Furthermore, protocols I, III and IV are vulnerable to the impersonation attack. Protocol III is also vulnerable to an offline dictionary attack. Excluding vulnerability of protocol I to the impersonation attack which has been implied in the standard, no attack has been reported on the protocols, and they are available in the IEEE 802.15.6 standard.

The impersonation attack is feasible because public keys are self-generated by involved parties, and they are not accompanied by digital certificates due to resource constraints in the nodes. Although this is not recommended in the standard, if one can use certified public keys, or we can have a lightweight PKI like the scheme proposed in \cite{dac2893}, this can prevent the impersonation attacks. However, all the protocols will still be vulnerable to the KCI attack. The KCI attack is a variant of the impersonation attack, and has been considered in the eCK security model \cite{eCK} for AKE protocols. Resilience to the KCI attack is an important security attribute for AKE protocols. If the private key of an entity $\mathcal{A}$ is compromised, an adversary $\mathcal{M}$ can impersonate $\mathcal{A}$ in one-factor authentication protocols. However, such compromise should not enable $\mathcal{M}$ to impersonate other honest entities in communication with $\mathcal{A}$. For the sake of briefness, we skip description of the KCI attack on protocols I, III and IV, because they are already vulnerable to the impersonation attack which is stronger than the KCI attack.

Forward secrecy is an important security attribute in AKE protocols. If an entity's private key has been compromised, it should not affect the security of session keys that have been established before the compromise. We have also the notion of \emph{Perfect Forward Secrecy} (PFS) which is a bit stronger than the forward secrecy. PFS means that the established session keys should remain secure even after compromising the private keys of all the entities that are involved in the protocol. We have the concept of weak-PFS which only allows a passive attack after compromise of all involved private keys.

Protocols I-IV use elliptic curve cryptography. Then, it is crucial to have the public key validation. Upon receiving an ephemeral or static public key, the recipient entity must validate it. Otherwise, the protocol would be vulnerable to further attacks. In description of protocols I-IV in the IEEE 802.15.6 standard, it has been mentioned that public keys should be validated. However, such validations are absent in corresponding figures in the standard.
If one implements the protocols according to the standard's figures, and does not consider public key validations, further security vulnerabilities will arise. There will be extra scenarios for impersonation attacks on the protocols. Furthermore, all the protocols will be vulnerable to an invalid-curve attack \cite{T10} whereby an attacker can extract the private key of another entity. We do not consider those extra vulnerabilities, and strongly recommend to validate public keys.

In the rest of this paper, $\mathcal{E}$ denotes the adversary in a passive attack, and $\mathcal{M}$ denotes the adversary in an active attack. The order of protocols and attacks does not imply any preference or importance. The numbering is according to the standard, and will be included in the $SSS$ during protocol executions.

\subsection{Protocol I}
\label{sec:4a}
Protocol I is an unauthenticated key exchange protocol. It is trivially vulnerable to an impersonation attack, but we consider it just for completeness of our security analysis. Such vulnerability has been implied in the standard only for this protocol, where the protocol is introduced as a protocol ``without the benefit of keeping third parties from launching impersonation attacks'' \cite{WBANstandard}.
Protocol I does not provide the forward secrecy.

\subsubsection{Impersonation attack:}
\label{sec:4a1}
Here is an impersonation attack on protocol I, in which $\mathcal{M}$ impersonates $\mathcal{A}$:
\begin{enumerate}
    \item[-] $\mathcal{M}$ selects a private key $SK_M$, and generates the corresponding public key as $PK_M = ({PK_M}_X , {PK_M}_Y) = SK_M \times G$. $\mathcal{M}$ selects a 128-bit random number $N_M$, and sends \{$ID_B || ID_A || SSS || AC || N_M || {PK_M}_X || {PK_M}_Y || XX$\} to $\mathcal{B}$.
    \item[-] $\mathcal{B}$ selects a 128-bit random number $N_B$, and sends \{$ID_A || ID_B || SSS || AC || N_B \\ || {PK_B}_X || {PK_B}_Y || XX$\} to $\mathcal{M}$.
    \item[-] $\mathcal{B}$ computes $DH_{Key} = X(SK_B \times PK_M)$, $T'_1 = RMB_{128}(DH_{Key})$, $T'_2 = CMAC(T'_1, ID_A || ID_B || N_M || N_B || SSS, 64)$, and $T'_3 = CMAC(T'_1, ID_B || ID_A \\|| N_B || N_M || SSS, 64)$. $\mathcal{B}$ sends \{$ID_A || ID_B || SSS || AC || N_B || {PK_B}_X || {PK_B}_Y || T'_2$\} to $\mathcal{M}$.
    \item[-] $\mathcal{M}$ computes $DH_{Key} = X(SK_M \times PK_B)$, $T_1 = RMB_{128}(DH_{Key})$, $T_2 = CMAC(T_1, ID_A || ID_B || N_M || N_B || SSS, 64)$, and $T_3 = CMAC(T_1, ID_B || ID_A \\|| N_B || N_M || SSS, 64)$. $\mathcal{M}$ sends \{$ID_B || ID_A || SSS || AC || N_M || {PK_M}_X || {PK_M}_Y$ $|| T_3$\} to $\mathcal{B}$. $\mathcal{M}$ computes $T_4 = LMB_{128}(DH_{Key})$, and generates the master key $MK = CMAC(T_4, N_M || N_B, 128)$.
    \item[-] $\mathcal{B}$ verifies that $T_3 = T'_3$, computes $T'_4 = LMB_{128}(DH_{Key})$, and generates the master key $MK = CMAC(T'_4, N_M || N_B, 128)$.
\end{enumerate}

$\mathcal{M}$ and $\mathcal{B}$ reach to the same $MK$ at the end. $\mathcal{M}$ could successfully impersonate $\mathcal{A}$.
A similar scenario for an impersonation attack can be written where $\mathcal{M}$ impersonates $\mathcal{B}$ in communication with $\mathcal{A}$.

\subsubsection{Lack of Forward Secrecy:}
\label{sec:4a3}
Here we show that Protocol I does not provide the forward secrecy, and then does not provide the weak-PFS or PFS:
\begin{enumerate}
    \item[-] Assume that $SK_B$ has been compromised. $\mathcal{E}$, that has eavesdropped and saved all the messages exchanged through previous runs of the protocol, knows $PK_A$, $N_A$ and $N_B$. $\mathcal{E}$ computes $DH_{Key} = X(SK_B \times PK_A)$, $T'_4 = LMB_{128}(DH_{Key})$, and obtains the established key $MK = CMAC(T'_4, N_A || N_B, 128)$.
    \item[-] If $SK_A$ has been compromised, $\mathcal{E}$ computes $DH_{Key} = X(SK_A \times PK_B)$, $T_4 = LMB_{128}(DH_{Key})$, and obtains $MK = CMAC(T_4, N_A || N_B, 128)$.
\end{enumerate}

\subsection{Protocol II}
\label{sec:4b}
Protocol II requires out-of-bank transfer of a node's public key to the hub. It is vulnerable to the KCI attack, and lacks the forward secrecy.

\subsubsection{Key Compromise Impersonation attack:}
\label{sec:4b1}
Protocol II is vulnerable to the KCI attack. Here is the attack scenario in which $\mathcal{M}$ has $SK_A$ and impersonates $\mathcal{B}$. As the public key of $\mathcal{B}$ is sent in clear, we can assume that $\mathcal{M}$ has obtained $PK_B$ by eavesdropping a previous protocol run.

\begin{enumerate}
    \item[-] $\mathcal{A}$ selects a 128-bit random number $N_A$, and sends \{$ID_B || ID_A || SSS || AC || N_A \\|| 0 || 0 || XX$\} to $\mathcal{B}$. $\mathcal{M}$ hijacks the session, and tries to impersonate $\mathcal{B}$.
    \item[-] $\mathcal{M}$ selects a 128-bit random number $N_M$, and sends \{$ID_A || ID_B || SSS || AC || N_M \\|| {PK_B}_X || {PK_B}_Y || XX$\} to $\mathcal{A}$.
    \item[-] $\mathcal{M}$ has $SK_A$. $\mathcal{M}$ computes $DH_{Key} = X(SK_A \times PK_B)$, $T'_1 = RMB_{128}(DH_{Key})$, $T'_2 = CMAC(T'_1, ID_A || ID_B || N_A || N_M || SSS, 64)$, and $T'_3 = CMAC(T'_1, ID_B||\\ID_A||N_M||N_A||SSS, 64)$. $\mathcal{M}$ sends \{$ID_A || ID_B || SSS || AC || N_M || {PK_B}_X || {PK_B}_Y \\|| T'_2$\} to $\mathcal{A}$.
    \item[-] $\mathcal{A}$ computes $DH_{Key} = X(SK_A \times PK_B)$, $T_1 = RMB_{128}(DH_{Key})$, and $T_2 = CMAC(T_1, ID_A || ID_B || N_A || N_M || SSS, 64)$. $\mathcal{A}$ verifies that $T_2 = T'_2$, and computes $T_3 = CMAC(T_1, ID_B || ID_A || N_M || N_A || SSS, 64)$. $\mathcal{A}$ sends \{$ID_B || ID_A || SSS || AC || N_A || 0 || 0 || T_3$\} to $\mathcal{M}$.
    \item[-] $\mathcal{A}$ computes $T_4 = LMB_{128}(DH_{Key})$, and generates the master key $MK = CMAC(T_4, N_A || N_M, 128)$.
    \item[-] $\mathcal{M}$ computes $T'_4 = LMB_{128}(DH_{Key})$, and generates the master key $MK = CMAC(T'_4, N_A || N_M, 128)$.
\end{enumerate}
$\mathcal{M}$ and $\mathcal{A}$ compute the same $MK$. $\mathcal{M}$ could successfully impersonate $\mathcal{B}$.

\subsubsection{Lack of Forward Secrecy:}
\label{sec:4b2}
Protocol II does not provide the forward secrecy. As it is assumed that $PK_A$ has been securely shared with $\mathcal{B}$, we just consider the case that $SK_A$ has been compromised. We show how $\mathcal{E}$ can extract previously established $MK$ from the eavesdropped messages which proves lack of forward secrecy and PFS:
As $PK_B$, $N_A$ and $N_B$ are sent in clear, we can assume that they are eavesdropped and saved by $\mathcal{E}$. $\mathcal{E}$ computes $DH_{Key} = X(SK_A \times PK_B)$, $T_4 = LMB_{128}(DH_{Key})$, and obtains $MK = CMAC(T_4, N_A || N_B, 128)$.

\subsection{Protocol III}
\label{sec:4c}
Protocol III is a PAKE protocol. It is vulnerable to impersonation and offline dictionary attacks. It does not provide the forward secrecy. %Protocol III is also vulnerable to the KCI attack, but we do not discuss it.

\subsubsection{Impersonation attack:}
\label{sec:4c4}
For performing an impersonation attack to Protocol III, $\mathcal{M}$ first eavesdrops messages between $\mathcal{A}$ and $\mathcal{B}$ in a protocol run. $\mathcal{M}$ then obtains $PK'_A$ and $PK_A$ from messages (1) and (4) of the protocol. $\mathcal{M}$ computes $Q' = PK_A - PK'_A$, and uses $Q'$ for an impersonation attack. Note that we have $Q' = Q(PW)$. Here is an impersonation attack on protocol III, in which $\mathcal{M}$ impersonates $\mathcal{A}$:

\begin{enumerate}
    \item[-] $\mathcal{M}$ selects a private key $SK_M$, and generates the corresponding public key as $PK_M = ({PK_M}_X , {PK_M}_Y) = SK_M \times G$. $\mathcal{M}$ computes $PK'_M = PK_M - Q'$. If $PK'_M = O$, $\mathcal{M}$ selects a new private and public key and continues the process until $PK'_M \neq O$. $\mathcal{M}$ selects a 128-bit random number $N_M$, and sends \{$ID_B || ID_A || SSS || AC || N_M || {PK'_M}_X || {PK'_M}_Y || XX$\} to $\mathcal{B}$.
    \item[-] $\mathcal{B}$ selects a 128-bit random number $N_B$, and sends \{$ID_A || ID_B || SSS || AC || N_B \\ || {PK_B}_X || {PK_B}_Y || XX$\} to $\mathcal{M}$.
    \item[-] $\mathcal{B}$ calculates $PK_M = PK'_M + Q(PW)$, and computes $DH_{Key} = X(SK_B \times PK_M)$, $T'_1 = RMB_{128}(DH_{Key})$, $T'_2 = CMAC(T'_1, ID_A || ID_B || N_M || N_B || SSS,\\ 64)$, and $T'_3 = CMAC(T'_1, ID_B || ID_A || N_B || N_M || SSS, 64)$. $\mathcal{B}$ sends \{$ID_A || ID_B \\ || SSS || AC || N_B || {PK_B}_X || {PK_B}_Y || T'_2$\} to $\mathcal{M}$.
    \item[-] $\mathcal{M}$ computes $DH_{Key} = X(SK_M \times PK_B)$, $T_1 = RMB_{128}(DH_{Key})$, $T_2 = CMAC(T_1, ID_A || ID_B || N_M || N_B || SSS, 64)$, and $T_3 = CMAC(T_1, ID_B || ID_A \\ || N_B || N_M || SSS, 64)$. $\mathcal{M}$ sends \{$ID_B || ID_A || SSS || AC || N_M || {PK_M}_X || {PK_M}_Y\\|| T_3$\} to $\mathcal{B}$. $\mathcal{M}$ computes $T_4 = LMB_{128}(DH_{Key})$, and generates the master key $MK = CMAC(T_4, N_M || N_B, 128)$.
    \item[-] $\mathcal{B}$ verifies that $T_3 = T'_3$, computes $T'_4 = LMB_{128}(DH_{Key})$, and generates the master key $MK = CMAC(T'_4, N_M || N_B, 128)$.
\end{enumerate}

$\mathcal{M}$ and $\mathcal{B}$ reach to the same $MK$ at the end. $\mathcal{M}$ could successfully impersonate $\mathcal{A}$. A similar scenario for an impersonation attack can be written where $\mathcal{M}$ impersonates $\mathcal{B}$ in communication with $\mathcal{A}$.

\subsubsection{Offline Dictionary attack:}
\label{sec:4c1}
Protocol III is a PAKE protocol with two-factor authentication. It requires both public keys and a shared password. For PAKE protocols, it is crucial to provide resilience to offline dictionary attacks. If an adversary could guess a password, he should not be able to verify his guess offline. For performing a dictionary attack on protocol III, it is sufficient that $\mathcal{E}$ eavesdrops messages between $\mathcal{A}$ and $\mathcal{B}$ in a protocol run. $\mathcal{E}$ then obtains $PK'_A$ and $PK_A$ from messages (1) and (4) of the protocol. $\mathcal{E}$ computes $PK_A - PK'_A = Q(PW) = (Q_X,Q_Y)$. As $Q_X = 2^{32} PW + M_X$ and $Q_X$ is known, this can be used as a verifier. $\mathcal{E}$ can then try probable passwords from a dictionary of most probable passwords, and check which password $PW$ will map to $Q_X$. This can be done very fast, and $\mathcal{E}$ can find the password $PW$ that is shared between $\mathcal{A}$ and $\mathcal{B}$.

\subsubsection{Lack of Forward Secrecy:}
\label{sec:4c3}
Protocol III does not provide the forward secrecy. As $PK_B$, $N_A$ and $N_B$ are sent in clear, we can assume that they are eavesdropped and saved by $\mathcal{E}$. If $SK_A$ is compromised, $\mathcal{E}$ computes $DH_{Key} = X(SK_A \times PK_B)$, $T_4 = LMB_{128}(DH_{Key})$, and obtains the master key $MK = CMAC(T_4, N_A || N_B, 128)$.

\subsection{Protocol IV}
\label{sec:4d}
Protocol IV is vulnerable to an impersonation attack, and lacks the forward secrecy. %Without public key

\subsubsection{Impersonation attack:}
\label{sec:4d1}
Here is an impersonation attack on protocol IV, in which $\mathcal{M}$ impersonates $\mathcal{A}$:

\begin{enumerate}
    \item[-] $\mathcal{M}$ selects a private key $SK_M$, and generates the corresponding public key as $PK_M = ({PK_M}_X , {PK_M}_Y) = SK_M \times G$. $\mathcal{M}$ selects a 128-bit random number $N_M$, and computes $W_M = CMAC(N_M, ID_A || ID_B || {PK_M}_X || {PK_M}_Y, 128)$. $\mathcal{M}$ sends \{$ID_B || ID_A || SSS || AC || W_M || {PK_M}_X || {PK_M}_Y || XX$\} to $\mathcal{B}$.
    \item[-] $\mathcal{B}$ selects a 128-bit random number $N_B$, and sends \{$ID_A || ID_B || SSS || AC || N_B \\ || {PK_B}_X || {PK_B}_Y || XX$\} to $\mathcal{M}$.
    \item[-] $\mathcal{B}$ computes $DH_{Key} = X(SK_B \times PK_M)$, $T'_1 = RMB_{128}(DH_{Key})$, $T'_2 = CMAC(T'_1, ID_A||ID_B||W_M||N_B||SSS, 64)$, and $T'_3 = CMAC(T'_1, ID_B||ID_A \\||N_B||W_M||SSS, 64)$. $\mathcal{B}$ sends \{$ID_A||ID_B||SSS||AC||N_B||{PK_B}_X||{PK_B}_Y||T'_2$\} to $\mathcal{M}$.
    \item[-] $\mathcal{M}$ computes $DH_{Key} = X(SK_M \times PK_B)$, $T_1 = RMB_{128}(DH_{Key})$, $T_2 = CMAC(T_1, ID_A||ID_B||W_M||N_B||SSS, 64)$, and $T_3 = CMAC(T_1, ID_B || ID_A \\|| N_B || W_M || SSS, 64)$. $\mathcal{M}$ sends \{$ID_B || ID_A || SSS || AC || N_M || {PK_M}_X || {PK_M}_Y \\|| T_3$\} to $\mathcal{B}$.
    \item[-] $Display_M$ will show $BS2DI(D)$ in which $D = CMAC(N_M||N_B, N_B||N_M||T_1, 16)$.
    \item[-] $\mathcal{B}$ verifies that $T_3 = T'_3$, computes $W'_M = CMAC(N_M, ID_A||ID_B||{PK_M}_X||\\{PK_M}_Y, 128)$, and verifies that $W_M = W'_M$. $Display_B$ will show $BS2DI(D')$ where $D' = CMAC(N_M||N_B, N_B||N_M||T_1, 16)$.
    \item[-] As $Display_M = Display_B$, $\mathcal{B}$ computes $T'_4 = LMB_{128}(DH_{Key})$ and $MK = CMAC(T'_4, N_M || N_B, 128)$. $\mathcal{M}$ computes $T_4 = LMB_{128}(DH_{Key})$ and $MK = CMAC(T_4, N_M || N_B, 128)$.
\end{enumerate}

$\mathcal{M}$ and $\mathcal{B}$ compute the same $MK$. $\mathcal{M}$ could successfully impersonate $\mathcal{A}$.
A similar scenario can be written for an impersonation attack where $\mathcal{M}$ impersonates $\mathcal{B}$ in communication with $\mathcal{A}$.

\subsubsection{Lack of Forward Secrecy:}
\label{sec:4d4}
Protocol IV does not provide the forward secrecy. As $PK_A$, $PK_B$, $N_A$ and $N_B$ are sent in clear, we can assume that they are eavesdropped and saved by $\mathcal{E}$.
\begin{enumerate}
    \item[-] If $SK_B$ has been compromised, $\mathcal{E}$ computes $DH_{Key} = X(SK_B \times PK_A)$, $T'_4 = LMB_{128}(DH_{Key})$, and obtains $MK = CMAC(T'_4, N_A || N_B, 128)$.
    \item[-] If $SK_A$ has been compromised, $\mathcal{E}$ computes $DH_{Key} = X(SK_A \times PK_B)$, $T_4 = LMB_{128}(DH_{Key})$, and obtains $MK = CMAC(T_4, N_A || N_B, 128)$.
\end{enumerate}

\section{Conclusion}
\label{sec:5}
The security of the IEEE 802.15.6 standard for WBAN \cite{WBANstandard} has been challenged in this paper. We analyzed the security of four key agreement protocols that are used for establishing a master key in the security association process of the standard. We showed that all four protocols have security problems. They are vulnerable to the KCI attack, and lack the forward secrecy. Furthermore, the first, third and fourth protocols are vulnerable to the impersonation attack. The third protocol is also vulnerable to the offline dictionary attack. Further attacks will be feasible if public keys are not validated. The standard aims to provide the confidentiality, authentication, integrity, privacy protection and replay defence. We could not find any indication that the security mechanisms in the standard provide privacy. However, our attacks show that the confidentiality and authentication are not achieved by the current security mechanisms in the standard.

\bibliographystyle{splncs}    % sorting: order of appearance
\bibliography{BAN}

\begin{thebibliography}{10}

\bibitem{T15b}
Toorani, M.:
\newblock On vulnerabilities of the security association in the {IEEE} 802.15.6
  standard.
\newblock In: Proceedings of Financial Cryptography and Data Security (FC'15)
  Workshops -- 1st Workshop on Wearable Security and Privacy (Wearable'15).
  (January 2015)

\bibitem{chen2011body}
Chen, M., Gonzalez, S., Vasilakos, A., Cao, H., Leung, V.C.:
\newblock Body area networks: A survey.
\newblock Mobile Networks and Applications \textbf{16}(2) (2011)  171--193

\bibitem{WBAN14survey}
Movassaghi, S., Abolhasan, M., Lipman, J., Smith, D., Jamalipour, A.:
\newblock Wireless body area networks: A survey.
\newblock Communications Surveys Tutorials, IEEE \textbf{16}(3) (2014)
  1658--1686

\bibitem{WBANstandard}
Association, T.I.S.:
\newblock {IEEE P802.15.6-2012 Standard for Wireless Body Area Networks}.
\newblock \url{http://standards.ieee.org/findstds/standard/802.15.6-2012.html}
  (2012)

\bibitem{HMQV}
Krawczyk, H.:
\newblock {HMQV}: A high-performance secure {Diffie-Hellman} protocol.
\newblock In: Advances in Cryptology--CRYPTO'05, Springer (2005)  546--566

\bibitem{menezes07another}
Menezes, A.:
\newblock Another look at {HMQV}.
\newblock Mathematical Cryptology JMC \textbf{1}(1) (2007)  47--64

\bibitem{T15a}
Toorani, M.:
\newblock On continuous after-the-fact leakage-resilient key exchange.
\newblock In: Proceedings of the 2nd Workshop on Cryptography and Security in
  Computing Systems (CS2'15), Amsterdam, Netherlands (January 2015)

\bibitem{T14a}
Toorani, M.:
\newblock {Cryptanalysis of a new protocol of wide use for email with perfect
  forward secrecy}.
\newblock Security and Communication Networks (2014)

\bibitem{BPR}
Bellare, M., Pointcheval, D., Rogaway, P.:
\newblock Authenticated key exchange secure against dictionary attacks.
\newblock In: Advances in Cryptology--EUROCRYPT'00, Springer (2000)  139--155

\bibitem{T09b}
Toorani, M., Beheshti, A.:
\newblock A directly public verifiable signcryption scheme based on elliptic
  curves.
\newblock In: Proceedings of the 14th IEEE Symposium on Computers and
  Communications (ISCC'09). (2009)  713--716

\bibitem{hankerson04guide}
Hankerson, D., Vanstone, S., Menezes, A.J.:
\newblock Guide to elliptic curve cryptography.
\newblock Springer (2004)

\bibitem{dac2893}
Misra, S., Goswami, S., Taneja, C., Mukherjee, A.:
\newblock Design and implementation analysis of a public key
  infrastructure-enabled security framework for {ZigBee} sensor networks.
\newblock International Journal of Communication Systems (2014)

\bibitem{eCK}
LaMacchia, B., Lauter, K., Mityagin, A.:
\newblock Stronger security of authenticated key exchange.
\newblock In: Provable Security, Springer (2007)  1--16

\bibitem{T10}
Toorani, M., Beheshti, A.:
\newblock Cryptanalysis of an elliptic curve-based signcryption scheme.
\newblock International Journal of Network Security \textbf{10}(1) (2010)
  51--56

\end{thebibliography}

\end{document}